# NASA'S NEXT GENERATION SPACE TELESCOPE
# VISITING A TIME WHEN GALAXIES WERE YOUNG

Bernard D. Seery (301) 286-5712, fax (301) 286-1670, bseery@hst.nasa.gov
Eric P. Smith (301) 286-8549, fax (301) 286-1753, ericsmith@stars.gsfc.nasa.gov
John C. Mather (301) 286-8720, fax (301) 286-1670, John.C.Mather.1@gsfc.nasa.gov

NASA Goddard Space Flight Center, Greenbelt, Maryland

Abstract
With the discovery of galaxies that existed when the universe was very young (approximately 5 percent of its current age), of planets not in our own solar system, and with the tantalizing evidence that the conditions for life may have existed within our solar system on planets or moons outside of the earth system, the past year has seen an explosion of interest in astronomy. In particular, a new era of exploration and understanding seems imminent, where the connection between the existence for the conditions of life will be connected to the origin of galaxies, stars and planets within the Universe. Who knows where this quest for knowledge will take us?

## 1.0 Science with NGST

Astronomy in the latter half of this century has made wondrous discoveries, expanded our understanding of the universe and opened humanity's vision beyond the visible portion of the electromagnetic spectrum. Our knowledge of how the cosmos was born and how many of its phenomena arise has grown exponentially in just one human lifetime. In spite of these great strides, fundamental questions are largely unanswered. In particular we have at best a rudimentary picture of how the primeval fireball cooled and formed the structures we are familiar with today, galaxies and stars. To further our understanding of the way our present universe formed following the Big Bang requires a new type of observatory with capabilities currently unavailable in either existing ground-based or space telescopes. NASA is currently studying such a telescope, currently known as the Next Generation Space Telescope (NGST) which is being designed to observe the first stars and galaxies in the Universe. This grand effort is embedded in NASA's Origins program, which seeks to investigate the origins of galaxies, stars, planets, and perhaps even life.

Recent discoveries by ground-based telescopes of galaxies at a redshift greater than 5 (when the universe was only one sixth it present size or about one tenth its present age) have profound implications for our understanding of the history of the universe. Observations by the Hubble Space Telescope (HST) have also revealed the large number of seemingly well formed galaxies whose redshifts are on the order of 2 to 3. For such galaxies to resemble present day systems they must have already been in existence for a few billion years. Other recent HST observations with the Near Infrared Camera/Multi-Object Spectrograph (NICMOS) have begun to show the interesting phenomena that can be studied in the near infrared at high angular resolution (~0.1 arcsec FWHM). These observations, and others like them have pointed the way to a regime where much progress is likely to be made in observational astronomy: high angular resolution infrared astronomy.

## 2.0 Mandate from the Astronomical Community

In 1993 the Association of Universities for Research in Astronomy (AURA) commissioned a study by a committee of 17 senior astronomers to study possible programs and outline astronomical goals for the first decade of 21st century space astronomy. This committee, chaired by Carnegie Observatories of Washington astronomer Alan Dressler, published its recommendations in 1996. The Dressler committee was comprised of astronomers whose experience spanned the infrared through ultraviolet portion of the spectrum, and whose areas of study ranged from planetary systems to cosmology. The committee actively solicited input from other astronomers and held public discussions at an American Astronomical Society meeting. Hence, the goals outlined by this report represent the desires of the infrared-optical-ultraviolet community for the coming decade. Their report titled "HST and Beyond: Exploration and the Search for Origins"[1] had three principal recommendations: (1) operate the Hubble Space Telescope beyond it nominal end of life (2003) by dramatically reducing the operations costs, (2) construct a near infrared optimized, large diameter (>4m) space telescope for the purpose of studying the birth of the first stars and galaxies, and (3) develop an infrared interferometer for the purpose of detecting planets around nearby stars. The second recommendation in this monograph has matured into the Next Generation Space Telescope concept.

## 3.0 NGST Core Mission

While the NGST will be a general use facility and consequently used for a wide variety of astronomical programs, it is important to remember the prime missions of the observatory will be those outlined in the Dressler committee report. These goals, which have percolated up from the astronomical community, are the study of the faint and distant universe. To better understand the role that NGST will play in the arsenal of astronomical telescopes it is instructive to consider the regions of space already mapped out by other space and ground-based telescopes. In Figure 1 we schematically show the current best understanding of the history of the evolution of structure in universe. The diagram indicates the regions from which we have already collected data. Signals from the first 300,000 years of the universe have been studied most completely by the Cosmic Background Explorer (COBE) satellite, and these results will be improved on by the soon to be flown on NASA's Microwave Anisotropy Probe (MAP) mission and European Space Agency's (ESA) Planck mission. These observatories study the light from the universe prior to the formation of the discrete structures we are all familiar with today, stars and galaxies. At the opposite end of time, nearer to the present, we have been able to study the universe with ground-based telescopes and space satellites like the Hubble Space Telescope. These instruments have probed the universe at ages from a few billion years old to today (12-15 billions years after the big bang). The very first stars and galaxies began to coalesce sometime after the first million years after the Big Bang and we know that galaxies already existed by approximately one billion years. The region of time betwwen these two epochs in the universe's history, a "Dark Zone" where no astronomical observations exist, is the primary target of the NGST.

Figure 1: Structure Formation History of the Universe

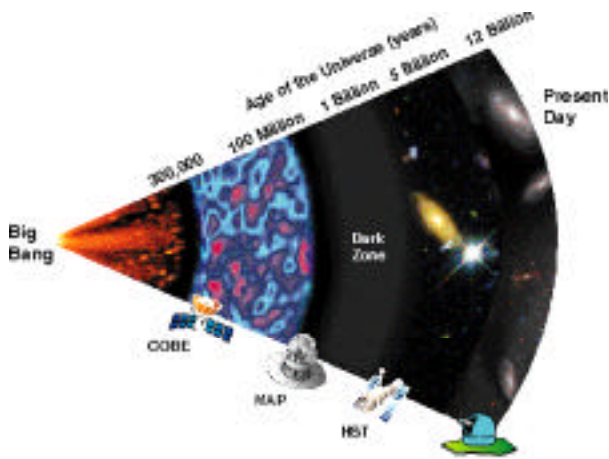

These first stars and galaxies are predicted to be extremely faint. Typical estimates for the brightness of primordial galaxies are in the 1-3 nanoJansky level (1 Jansky = $10^{-26}$ Watts $m^{-2}$ $Hz^{-1}$). The light from stars and galaxies in the dark zone will be redshifted to ~6-7 times their rest frame wavelength. Visible light will appear at approximately 2 microns. In the near infrared, ground-based observations are severely affected by atmospheric absorption and emission and the thermal radiation of the telescope optics. With angular resolution comparable or superior to that obtained under optimum conditions with adaptive optics, a large passively cooled space telescope will enjoy 1000 times less background and will be capable of extending the sensitivity of HST to wavelengths of 3-5µm, depending on the effective aperture.

The angular sizes of early galaxies are on the order of 0.2-0.5 arcseconds in diameter. While the angular resolution required for these studies is achievable with the Hubble Space Telescope, and, using active or adaptive optics, with some ground-based observatories, neither can combine the required flux level sensitivity and angular resolution over wide fields of view that the NGST can achieve.

4.0 Unique Capabilities

The demanding performance parameters discussed above drive the NGST into a unique portion of the observational phase space defined by sensitivity, angular resolution, and wavelength coverage. Each of the three parameters is critical to achieveing the goals of the NGST. Infrared wavelength coverage is required because the light from the distant objects is redshifted into these wavelengths. High angular resolution is critical because the objects we seek to discover are very small. Finally, high sensitivity is vital because the signals from these distant objects are very weak. Below in Figure 2 we show the relative performance (quantified as field-of-view times observing speed) of an 8m diameter, passively cooled NGST to other astronomical observatories. Performance for NGST is normalized to unity. NGST will provide orders of magnitude increases in speed capabilities, and be open to regions of the spectrum blocked by the atmosphere for ground-based observatories (e.g. the Gemini and VLT telescopes). Its high angular resolution will enable it to follow-up the lower resolution observations of the Space Infrared Telescope Facility (SIRTF) as well as discover new phenomena in the 3-20µm wavelength band

5.0 A General User Facility

It is clearly impossible to say exactly what science problems will be interesting in ten years, but there are some areas where we can generally predict NGST will yield unique observations. These areas stem primarily from the HST & Beyond report and are expressed in practical terms in the NGST Design Reference Mission (DRM)[2]. In particular, the HST & Beyond program of observations form the "floor goals" for NGST. The ultimate design of the NGST must be capable of addressing at least these studies. Hence, The DRM contains a suggested list of astronomical programs that define the envelope of performance requirements of the observatory. The DRM is comprised of potential astronomical targets, their expected physical properties (number density and brightness), and desired observation modes (wavelength band, spectral resolution, number revisits, etc.). In Table 1 below we present a top-level summary of the NGST DRM for both imaging and spectroscopic observations.

Figure 2: NGST Capabilities

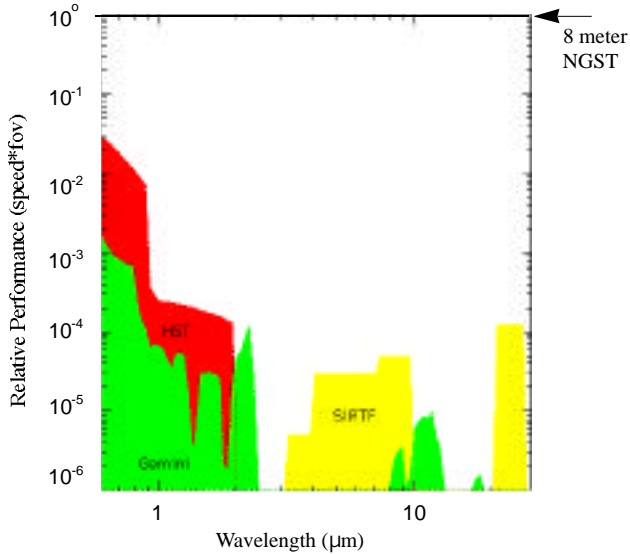

For each of the proposed studies in the DRM the NGST will far exceed the capabilities of other observatories. Questions about the birth and death of stars, the formation of galaxies, and the build up of heavy elements in the universe will be answerable with observations from this powerful observatory. These DRM programs, and other programs proposed by guest observers will, like the HST before it, push our understanding of the universe to new depths.

6.0  Programmatic Challenges

The NGST will be developed by NASA in an environment characterized both by advanced technology and cost constraints. The NGST goal is to develop order of magnitude increased scientific performance for an affordable cost to the public. It is clear from the excitement associated the Mars Pathfinder and Hubble Space Telescope missions that the public is NASA's ultimate customer. It was felt that NGST development should be cost-capped at $500M in year 1996 dollars, and that launch costs and subsequent 10 year scientific and mission operations costs should not exceed $400M (96), thus, for a total life cycle cost to the taxpayer of $900M (96).

So how can we achieve these challenging cost goals, given the reality of the cost of other similar, perhaps even less ambitious missions? NASA believes that the answer may lie with innovative new technologies. Technologies which may allow mission designers to achieve these striking levels of performance at reduced cost. In the past, technology was a way to fix problems in spacecraft design. Today's paradigm is one of technology as a mission enabling element, not a problem solver. However, to achieve this reality, a significant emphasis on developing these breakthrough technologies must be present up front in the mission design process. NGST is perhaps the first large astronomical mission with such an emphasis.

Technical or performance breakthroughs in lightweight optics and passive cooling are important, but NGST cannot be built to cost based on this alone. Success will come only if the methodology, for spacecraft development is re-engineered as well. New contracting methods, referred to

Table 1: NGST Design Reference Mission

| Program | Wavelength band (microns) | Typical target fluxes (nanoJanskys)* |
|---|---|---|
| COSMIC DISTANCES (supernovae studies) | 1-5 | ~5 |
| PRIMEVAL GALAXIES | 1-5 | 1-100 |
| GRAVITATIONAL LENSING | 1-5 | ~300 |
| CHEMICAL EVOLUTION (distant Inter Stellar Medium (ISM)) | 0.5-5 | ~5000 |
| ACTIVE GALAXIES | 1-20 | ~300 |
| NORMAL GALAXIES | 0.3-20 | $2 \times 10^5$ |
| STELLAR DEATH | 10-20 | $5 \times 10^5$ |
| LOCAL STELLAR POPULATIONS | 0.5-2 | 3-12 |
| LOCAL ISM | 5-20 | $1 \times 10^6$ |
| BROWN DWARFS | 5-20 | 6000 |
| CIRCUMSTELLAR GAS/DUST | 5-20 | 6000 |
| SOLAR SYSTEM STUDIES | 1-20 | $14-4 \times 10^4$ |

*(1 nanoJansky = $10^{-34}$ Watts m$^{-2}$ Hz$^{-1}$)

as Performance Based, which incentivize the development contractor to optimize the observatory for on orbit performance, based on pre-determined metrics, are being adopted for NGST[3]. Additionally, significant emphasis is being placed on detailed analytical models and computer simulations of all aspects of the design, including the costly operations and data reduction phases of the mission. Hardware testbeds will be employed where necessary to validate these models, with the ultimate goal being to fully simulate the end-to-end system performance.

7.0 Derived Requirements

Responding to our strawman science program summarized in Table 1, the NGST scientists and engineers studied the role of the main instrumental parameters on the scientific capability of the observatory, using the completion rate of the DRM program over a given mission lifetime as a figure of merit. We established a science floor; that is, the minimum top level performance criteria to satisfy the recommendations of the HST and Beyond committee. Then we set out a series of "stretch" goals, ones that would be highly desirable from a scientific standpoint, but that must not be cost drivers. These goals are summarized in Table 2, along with the relevant HST parameters for comparison purposes, since as the successor to HST, it is often compared against it.

The NGST science program is driven by sensitivity in natural background-limited conditions, not simply by angular resolution. As a result, the primary mirror aperture configuration should be as compact as possible, with a full circular aperture preferred. In fact, the diameter of the collecting aperture must exceed 6 meters in order to adequately complete the core and most of the complementary science programs in a 5 year mission. To some extent, telescope diameter can be traded off against instrument field of view, however. The instruments, typically less expensive than telescopes, were baselined with as large a field as practical given the limits defined by aberration theory and mechanical packaging considerations.

NGST is optimized for the near infrared because the radiation emitted by early universe objects is "redshifted" into the wavelength range in the neighborhood of 1-5 micrometers. The temperature of the telescope and instrument optics must be maintained below 100K so that the optics emission is negligible in the near infrared (NIR) wavelength bands of interest. To achieve background-limited performance for the full gamut of the mid infrared (MIR) science program, the telescope optics must be less than 35K. A summary of the desired characteristics of the observatory are shown in Table 3.

Table 3 Desired Scientific Requirements

| Item | Required Performance |
|---|---|
| Angular resolution | <60 milliarcsec at 2mm |
| Limiting magnitude | up to 32 ABm |
| Spectral range | 1 to 5 mm (0.6 to 26mm goal) |
| Spectral resolution | up to 4000 |
| Field of view ( simultaneous imaging) | 4' x 4' |
| Instrumental background | <zodiacal light at 1-5mm |
| Optics temperature | <60 K (30 K goal) |
| Detector dark current | < 0.02 electrons sec-1pixel-1 |
| Instantaneous sky coverage | >20% available |
| Mission sky coverage | 100% available |
| Mission Lifetime (yr) | > 10 |

Table 2 Floor & Stretch Goals

| Parameter | HST | NGST Science Floor | NGST Science Goals |
|---|---|---|---|
| Wavelength Range | Ly - 2µm ~ | Near Infrared | 0.5 - 30µm |
| Angular Resolution | Diffraction-Limited at 0.55µm | Diffraction-Limited at 2µm | Diffraction-Limited at 0.55µm |
| Aperture Diameter | 2.4m | >4m | >8m |
| Sensitivity | Instrument-Limited (NICMOS) | Zodi-Limited at 1 AU | Exo-Zodiacal background-Limited |
| Lifetime | 15 years | >5 years | 10 years |
| Instruments | WFPC2, STIS, NICMOS, FOC, FGS | Wide Field Camera/ Spectrometer | Add visible MIR Camera/ Spectrometer and Coronograph |
| InstrumentsWFPC2, STIS, | Wide Field Camera/ NICMOS, FOC, FGS | Add visible Spectrometer | MIR Camera/ Spectrometer and Coronograph |

8.0 NGST Mission Concept

The early phases of the NGST concept studies have produced at least four candidate architecture from our Industry partners, LMMS, TRW, and Ball, spanning the range of requirements as well a cost. These are conceptually illustrated in figure 3 relative to where they rank in the requirement and cost categories.

A reference design concept, known as the "Yardstick Mission," was established early on by the NASA team to serve as a fiducial against which all other designs could be "measured." Specifically, it provides both a feasibility proof and a cost analysis tool for evaluating current and future industry designs.

The design premise was to analytically demonstrate that a primary mirror could be made larger than its stowed dimension dictated by the launch vehicle fairing. Since launch cost is included in the mission lifestyle cost cap, it is obvious that the lower the launch cost, or the smaller the vehicle, the more resources are available for science capability. Segmented deployable structures, pioneered by the DoD in the eighties, seemed to offer an answer that was robust to the vagaries of the launch vehicle industry. This approach, when applied to an optical system, is enabled by the technology and methodology of active optics.

The "Yardstick Mission Concept" uses a segmented deployable 8 meter diameter telescope optimized for the near infrared region (2-5 microns), but with instruments capable of observing from the visible all the way to 30 microns. The observatory is radiatively cooled to about 30 K and would be launched on an Atlas IIAS or equivalent rocket to the Lagrange Point L2. The main characteristics of the observatory are shown in Table 4.

Figure 4 shows the overall view of the observatory and its main components. It is composed of the Optical Telescope Assembly, Science Instruments Module, and the Space Support Module.

Figure 3 Concept Suite Spans Requirements Phase Space

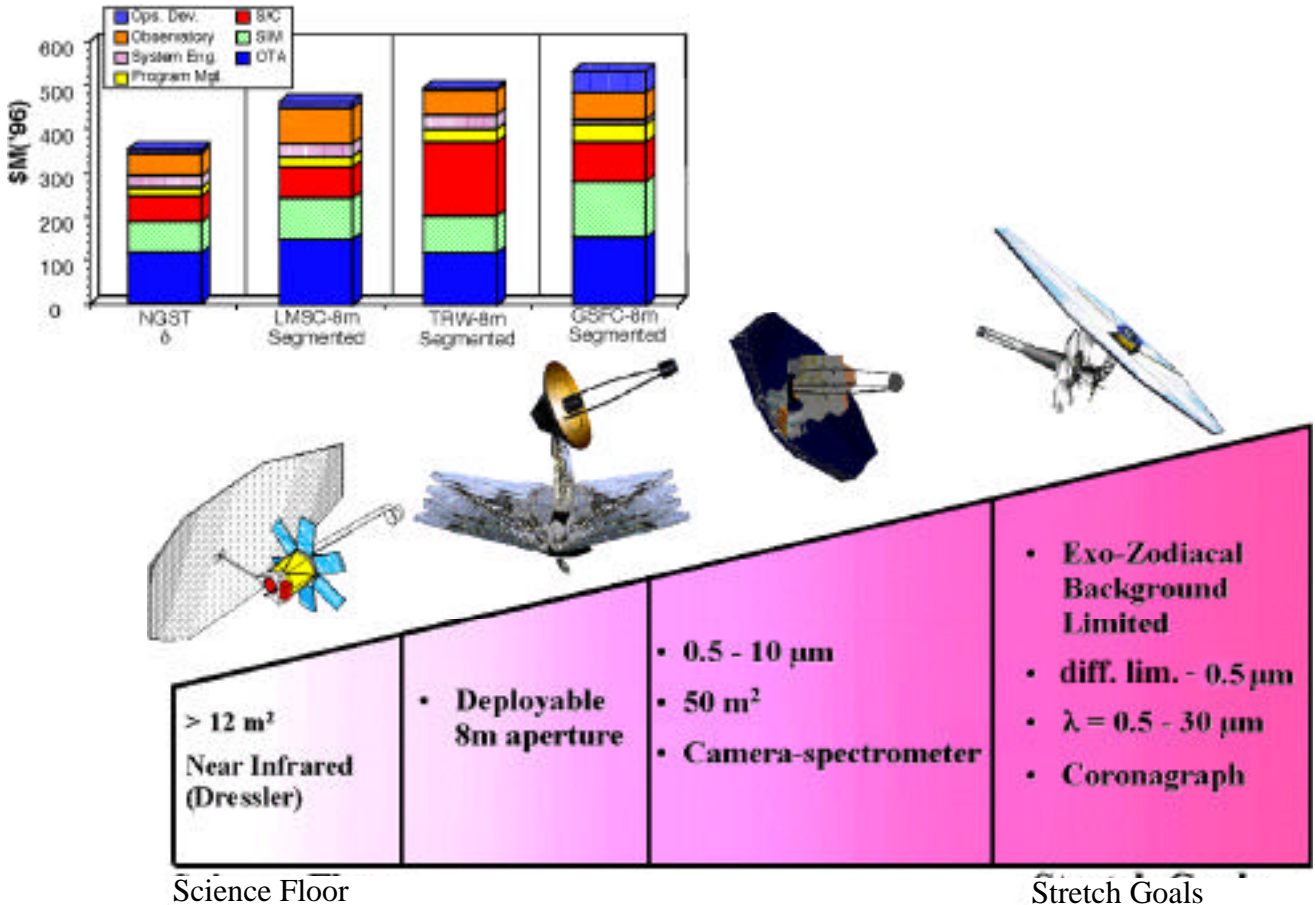

Figure 4 Artist view of the yardstick Observatory and its main components

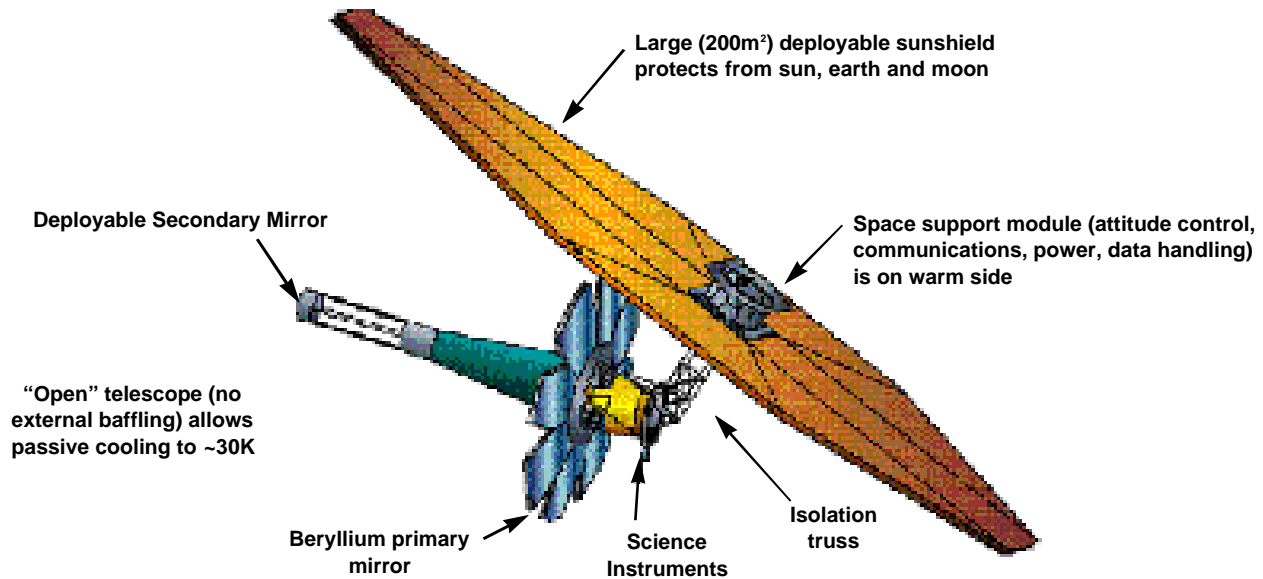

Table 4 - NGST Yardstick architecture main characteristics

| Item | Value |
| --- | --- |
| Scientific Performance | |
|    Wavelength | 0.6 to 30µm |
|    Aperture | 8 meter, quasi-filled |
|    Sensitivity | 4 nJy in 10,000s at 2µm, S/N=10, BP20% |
|    Resolution | 0.050" (diffraction limited at 2µm |
|    Science Instrument | cameras, multiobject spectrograph |
|    Field | NIR: 4'x4' (camera) |
| | 3'x3' (spectrograph) |
| | MIR: 2'x2' (camera) |
|    Sky coverage | Yearly: full sky |
| | Instantaneous: 17% |
| Technical features | |
|    Optics configuration | 3-mirror anastigmat with accessible exit pupil |
|    Aperture diameter | 8 meters OD, 7.2m effective dia |
|    Wavefront control | image-based wavefront sensing with 5 Degrees of Freedom mirror actuation + Deformable mirror |
|    Optics temperature | <70 K (50 K nominal) |
|    Mirror material | lightweight beryllium |
|    System f ratio | f/24 |
|    Fine pointing | 10 mas |
|    Data rate | 1.6 mbps |
| Mission aspects | |
|    Mass | <3300 kg |
|    Spacecraft pointing | 2"rms |
|    Power | 800W |
|    Mission Lifetime | 5 years nominal - 10 years goal |
|    Orbit | Sun-earth Lagrange 2 Halo orbit |
|    Launcher | Atlas lll a |
| Programmatic aspects | |
|    Cost of manufacture | $564M (1996 $) |
|    Development duration | 48 months |
|    Launch date | June 2007 |

Figure 5 NGST Technology Roadmap

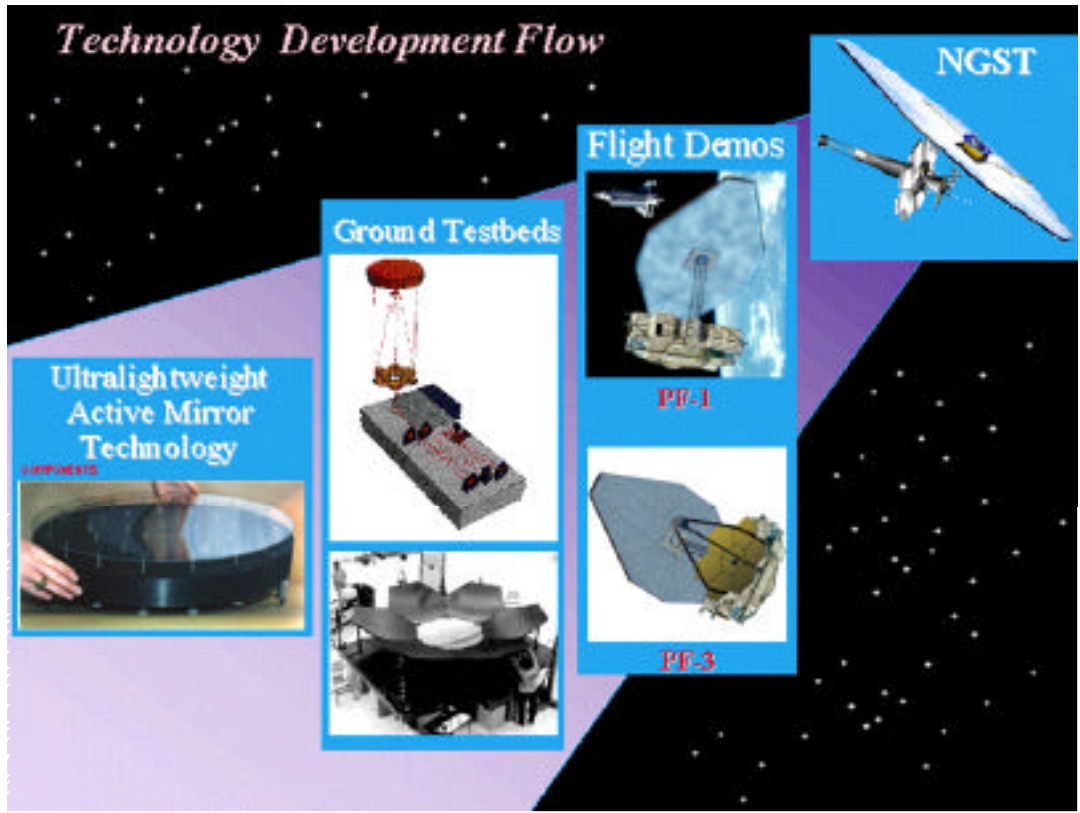

## 9.0 Technology Roadmap

A prime contractor will be chosen to begin detailed design and development of the observatory space and ground segments sometime after the year 2001. Until that time, NASA intends to develop the technology necessary to insure that a high performance NGST can be manufactured for a cost of $500M. Breakthroughs in technology are needed both in terms of performance at a lower cost and reduced development schedule. Much work has already been accomplished early in the program and the groundwork has been laid to further develop innovative new mirrors and precision deployable structures.

The NGSTtechnology development roadmap for readying the enabling technologies is illustrated in Figure 5.

The figure depicts a strategy that begins with component development, evolves to subsystem and system level testbeds, and culminates in selected flight validation experiments.

## 10.0 Active Optics

The most critical and probably the most challenging of component technologies for NGSTis the ultra-lightweight mirror technology. Systems studies have shown that the primary mirror assembly is the chief driver of system mass. Therefore, the chief emphasis in our technology program is here. Figure 6 is a graph of areal density, in units of kilograms per square meter, as a function of calendar year since the completion of the HST mirror at 250 kg/m2. In the intervening period since HST and prior to the onset of NGST in 1996, the Department of Defense has been the principal developer of active optics technology. In fact, NGST is in some way made feasible by that DoD investment in the technology.

The NGST goal for areal density is 15 kg/m2, about a factor of 2 improvement over the current state of the art. This density includes the thin membrane reflector, figure actuators, and backing structure, including wires. There are four mirror development activities funded by the NGST Project. Under the NGST Mirror System Demonstrator (NMSD) contract, the University of Arizona and Composite Optics Incorporated (COI) are developing ambitious 2 meter-class active optics using a thin glass membrane and a glass-composite sandwich, respectively. The third development effort is an ultra-lightweight Beryllium mirror at the half meter diameter scale being developed by Ball Aerospace. A fourth activity is the demonstration of fine polishing of a carbon reinforced silicon carbide mirror blank manufactured by IABG in Germany.

Figure 6 Areal Density

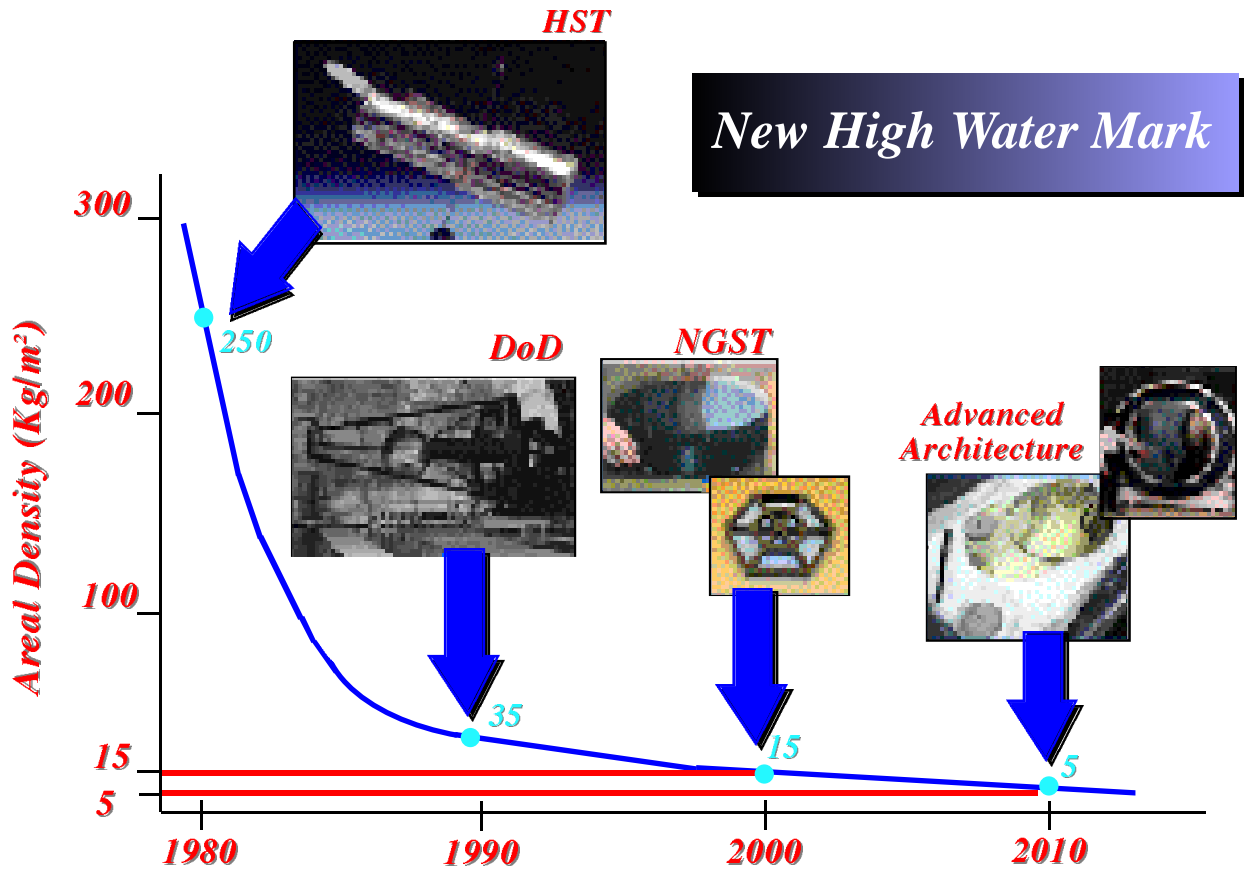

## 11.0 University of Arizona

The University of Arizona team includes Thermo Trex Corporation (TTC) who has responsibility for the cryogenic actuator development and the mirror figure control system, Composite Optics Inc. (COI) who has the responsibility for the load spreader/flexure and reaction structure manufacture, and Lockheed Martin who has responsibility for reaction structural design and system testing. UoA has overall responsibility for system integration and for the grinding and polishing of the mirror face sheet. The UoA mirror is hexagonal, 2 meters in diameter (point to point). It has a spherical figure with a radius of curvature of 20m. It consists of a 2 µm thick borosilicate (Schott Borofloat) glass membrane attached to 200 actuators via 9 point loadspreader/flexures. The actuators are placed on a composite reaction structure at 7cm intervals. The system has an areal density of 13 kg per square meter. It is to be diffraction limited at 2 microns. The basic concept, shown in Figure 7, is that the actuator spacing, resolution and throw are such that is will be possible to correct for mirror manufacturing errors, cryogenic distortion and gravitational compensation during testing and in space - zero-g release cryogenic cool down and temperature gradients and drift.

Figure 7 UoA Mirror Concept

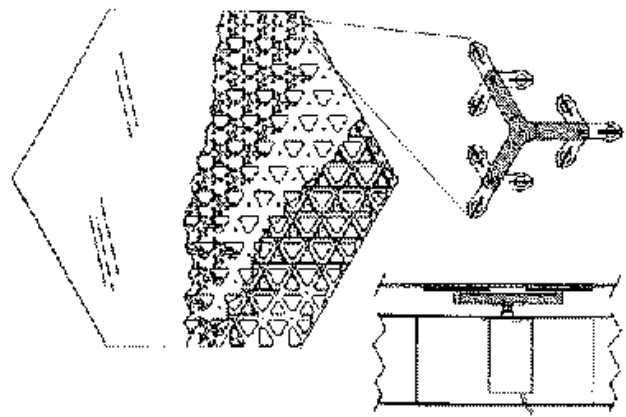

## 12.0 Composite Optics, Inc.

The COI mirror is a hybrid glass composite mirror. It is 1.6 meters in diameter and is circular in shape with one flat side. The mirror consists of a 3.2 mm zerodur (Schott) glass face sheet bonded to a composite core with a composite back sheet. The mirror is mounted via flexures and a central force actuator for radius curvature adjustment to a composite reaction structure. The mirror has a spherical figure with a 15 meter radius of curvature and is also designed to be diffraction limited at 2 microns and have an areal density of less than 15 kg per square meter. The overall concept is shown in Figure 8.

## 13.0 Ball Aerospace & Technology, Inc.

The Subscale Beryllium Mirror Demonstrator (SBMD) configuration being developed by Ball Aerospace is a 532 mm diameter light-weighted beryllium mirror with one flat edge and a corresponding solid beryllium backplane connected by a series of actuators and loadspreaders. Presently the design is for three tip/tilt/piston actuators and one radius of curvature correction actuator in the center. The SBMD mirror and backplane will be fabricated from a single spherical powdered Beryllium boule supplied by Brush Wellman. The SBMD proposed mirror design has an areal density of approximately 12.3Kg/m2, consisting of a 2.5 mm facesheet, 1.0 mm thick ribs, with a triangular cell open area design using 50 mm inscribed circle spacing.

## 14.0 IABG CSiC

A contract is in place with IABG to produce a 0.5m carbon reinforced Silicon Carbide mirror blank which will then be provided as government furnished equipment on a contract which will be open bid to grind and polish the blank. Coupons have been received from IABG which will be provided to optics houses that are interested in grinding and polishing the 0.5m blank. These coupons will be used to gain experience in grinding and polishing the CSiC material.

## 15.0 Wavefront Sensing & Control Methodology

Technology for alignment, phasing and wavefront sensing and control will be developed and validated by the Developmental Cryogenic Active Telescope Testbed (DCATT). A concept for a layered control approach utilizing image based sensing, as well as the software to implement it, is under development, and shown schematically in figure 9.

DCATT will implement the approach in subscale hardware by development of an actively controlled 1 m class telescope with a segmented primary. The primary segments will be actuated and controlled in 6 Degrees of Freedom (DOF) and the secondary will also be controlled in 6 DOF. An active optics bench will include relay optics, a deformable mirror, a fast steering mirror and a coarse and fine wavefront sensor. A stimulus and scoring interferometer will complete the testbed. Operation is planned in three phases. Phase 1 will include room temperature operation with mostly off the shelf components to investigate and validate the fundamental optical control scheme. In Phase 2, the telescope (with its mostly off the shelf components) will be re-located to a cryogenic chamber during operation. The active optics bench will remain outside at ambient temperature. The principal goal of this phase is to evaluate performance of the cold telescope relative to the predictions of the NGST Integrated Models. Finally, in Phase 3, the telescope will be retro-fitted with flight-like components. The active optics bench will remain at ambient temperature. These efforts will investigate and evaluate techniques for coarse and fine wavefront sensing, coarse alignment of the primary mirror segments and the secondary mirror, phasing of the primary mirror segments, wavefront correction with the deformable mirror, and image stabilization with the fast steering mirror as well as thermal effects on the telescope.

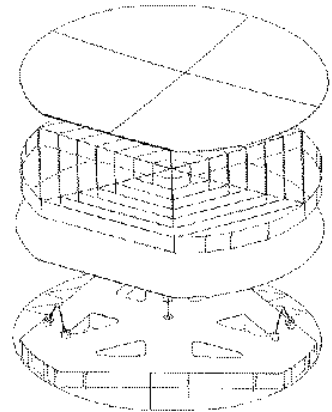

Figure 8 Composite Optics

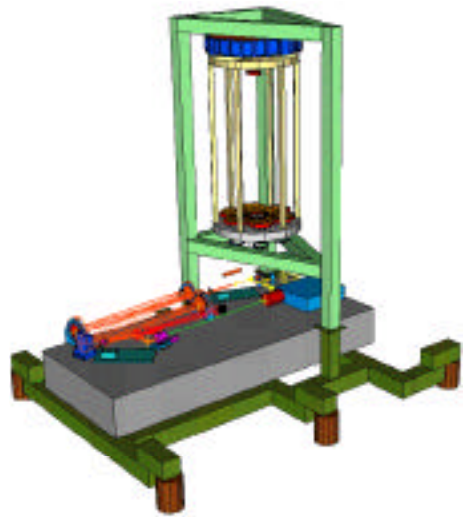

Figure 9 DCATT Testbed Developmental Cryogenic Active Telescope Testbed

## 16.0 Ground Testbeds and Flight Experiments

In addition to the DCATT Testbed, an NGST Flight System Testbed is planned to be a nearly full scale, fully functional, "flight like" model of the NGST. It will be built at the system contractors facility after the down-selection to a single contractor. It will likely be an outgrowth of one or more testbeds built during the earlier technology development activities, however, it will utilize the architecture and technologies that have been selected for the flight build and will validate many aspects of the system level performance. It is anticipated that successful development and operation of this testbed during phase B will be a key element in proving readiness to move to the implementation phase of NGST. It is also expected to be useful during the flight build, integration and test phase and serve much of the role traditionally served by an "engineering model".

The first Pathfinder flight experiment, the Inflatable Sunshade In Space, or ISIS, is planned to be a low-cost flight demonstration and validation of an inflation deployed sunshade for NGST. Current plans call for an approximately half scale version of the shade to be flown on a shuttle deployed, free-flying spacecraft such as the GSFC Spartan. The duration of the flight would be expected to span one shuttle mission. The goals of the experiment are to test out suitable membrane materials, demonstrate a controlled inflation sequence, demonstrate rigidization of the support structure, measure flight dynamics, measure thermal performance and compare results to predictions of the NGST system models. Many elements of the shade performance simply cannot be practically tested on the ground and this flight is required to understand the risk of using this new technology for a large sunshade for NGST.

The final Pathfinder flight experiment, termed Pathfinder 3, is planned to demonstrate deployable optics in the space environment. This is perhaps the most important technology demonstration to validate a deployable telescope as a practical approach to large space telescopes and reduce the risk for using large deployable optics for space telescopes. The goal is to build, launch and operate a segmented, deployable 2.2m diameter aperture telescope and demonstrate diffraction limited imaging of a bright star or other celestial object. The flight will be on a long duration version of the Spartan spacecraft and will span at least two shuttle flights (several months). The flight will be near the end of the Formulation Phase of NGST and; along with the Flight System Testbed and other accomplishments, will provide evidence that the NGST team is ready to proceed to the Implementation Phase. Artist conceptions of the two Pathfinder flights are shown in Figure 5.

To learn about the latest developments in the NGST Project, visit http://ngst.gsfc.nasa.gov/on the World Wide Web.

## 17.0 Summary

NASA's Next Generation Space Telescope (NGST) will see the dark universe light up with stars. It will observe the formation of the first galaxies and supernovae, and trace the formation of the heavy elements from the primordial hydrogen and helium of the earliest epoch. Technologically, it will re-define the state of the art in lightweight optics and detectors. Powerful onboard computers will be used to effectively rigidize the large, flexible structures and expansive mirror assemblies. To solve such immense challenges, strategic partnerships have been formed between GSFC, Space Telescope Science Institute, and key NASACenters, such as with the Marshall Space Flight Center and JPL, and between NASA and Industrial Aerospace firms such as Lockheed, TRW and Ball, to develop breakthrough technologies which will enable the type of 'order of magnitude' performance improvements needed to achieve our scientific goals, but at an affordable cost to the public. International participation is on the rise, and will further enhance NGST's mission, while strategic partnerships with the Air Force Research Labs is enriching the pipeline of revolutionary new technologies available to NGST. New methodologies, such as the Intelligent Synthesis Environment (ISE) provide the needed cultural change in the engineering and mission synthesis environment. These high performance computing and virtual product development tools are already being employed in the early trade and concept study phases of NGST by collaborative teams from NASA, the European Space Agency (ESA) and the Canadian Space Agency (CSA).

## 19.0 Acknowledgements

The authors wish to thank the many men and women working on NGST, both here and abroad. While this article was written largely with reference to the NASA "Yardstick" mission concept, we wish to acknowledge the work done by our Industry Partners, Lockheed Martin, TRW, and Ball Areospace. In conclusion, we would like to thank Dr. John Campbell and Ms. Mary Kicza at NASA/GSFC for their programmtic support, and Dr. Ed Weiler, Dr. Harley Thronson, and Mr. Rick Howard from NASAHeadquarters from sponsoring this exciting project.